\documentstyle[prl,aps,psfig,twocolumn]{revtex} 

\begin{document}
\twocolumn[\hsize\textwidth\columnwidth\hsize\csname
@twocolumnfalse\endcsname
\title{RANDOM DYNAMICAL SYSTEMS, ENTROPIES AND INFORMATION}
\author{Maurizio Serva}
\address{Istituto Nazionale di Fisica della Materia
and Dipartimento di Matematica}
\address{Universit\`a dell'Aquila, I-67010 Coppito, L'Aquila, Italy}
\bigskip

\date{\today}

\maketitle

\begin{abstract}
Prediction of events is the challenge in many different disciplines, 
from meteorology to finance; the more this task is difficult, 
the more a system is {\it complex}. Nevertheless, even according
to this restricted definition, a general consensus on what should be
the correct indicator for complexity is still not reached. 
In particular, this characterization is still lacking for systems 
whose time evolution is influenced by factors which are 
not under control and appear as random parameters or random noise.
We show in this paper how to find the correct indicators for complexity  
in the information theory context. The crucial point is
that the answer is twofold depending on the fact that the random 
parameters are measurable or not. The content of this apparently
trivial observation has been often ignored in literature leading to
paradoxical results. 
Predictability is obviously larger when the random parameters
are measurable, nevertheless, in the contrary
case, predictability improves
when the unknown random parameters are time correlated.
\end{abstract}
]
\narrowtext

In a number of systems the dynamics is influenced by uncontrolled  
parameters which are intrinsically random or cannot be predicted with 
necessary precision. 
The evolution of a system of this type is described in the framework
of random dynamical systems, word which indicates 
in the present paper also dynamical systems with noise.

A dynamical system can be eventually studied by means of
the associated symbolic dynamics, which, in this case,
correspond to a stochastic process with random conditional probabilities, 
i.e. probabilities which depend on the same stochastic parameters. 

The obvious thing is that the possibility of forecasting 
the future evolution strongly depends on the possibility of
measuring the parameters. The same model will have a different complexity
(predictability) according to the fact that the measure is feasible or not.
Even if the content this observation appears trivial it is often ignored
in literature.
For example, most frequently it is
used a definition of complexity which considers
the separation of nearby 
trajectories~\cite{CFH,BJS,MT} under the same realization 
of the noise. This definition implicitly assumes that
the realization is known and should not be used when
the contrary happens, has often it is.
For example, the  phenomenon of  
noise induced order~\cite{MT} should be not
considered a reduction of complexity
when the random disturbance is non measurable.

A better characterization of complexity for dynamical systems
with unmeasurable randomness has been recently found out for
many physically relevant cases~\cite{PSV,LPV,LSV,LPPV}.

In this paper we show how to find proper indicators of complexity
in the two cases of measurable (accessible information)
and non measurable (inaccessible information) randomness. 
We also show, with an example, that in case the stochastic 
parameters have memory, part of the inaccessible information is 
encoded in the dynamics of the system and can be recovered. 
In other words, the gap between the two indicators of 
complexity reduces when the random parameters are time correlated.

Let us start with some basic definition for the non random
case also in order to establish the notation. 
Let us assume that the state of the system
is identified by the vector $y(t)$
which evolves as a deterministic dynamical system
according to $y(t+1)=f(y(t))$.
The corresponding phase space can be partitioned 
in regions indexed by a symbol $x$. 
The associated symbolic dynamics $..,x(1),x(2),.., x(t),..$ 
is the realization of a stochastic process with memory,
i.e. the probability that the system is in $x(t+1)$  
depends on its past history $..,x(1),x(2),..,x(t)$.

The best characterization of predictability
of a process with memory can been found in the information 
theory context, and it is the Shannon entropy~\cite{SW}.
Assume that $\rho ({\bf x}_n)$ is the probability 
that the sequence of $n$ symbols $x(t+1),.., x(t+n)$ 
equals ${\bf x}_n \equiv x_1,..,x_n$ , in this case

\begin{equation}
H_n = - \sum_{ \{{\bf x_n}\}}
\rho ({\bf x}_n) \;\log \rho ({\bf x}_n)
\label{H_n}
\end{equation}
is the entropy of the sequence. Then, the entropy rate
$h_n =H_{n+1}-H_n$ ($n \ge 0,\, H_0 \equiv 0$) measures 
the average information 
contained in  $n$ steps of the process. In fact, the probability that
the system is in $x_{n+1}$ if it was in $x_1,..,x_{n}$
is $ \rho (x_{n+1} | {\bf x}_n)=  \rho ({\bf x}_{n+1}) / 
\rho ({ \bf x}_n)$
and one has

\begin{equation}
h_n = \sum_{ \{{\bf x}_n\}}
\rho({\bf x}_n) \; e_n({\bf x}_n)
\label{h_n}
\end{equation}
where

\begin{equation}
e_n({\bf x}_n)
= - \sum_{ \{x_{n+1}\}}
 \rho (x_{n+1} |{\bf x}_n) 
\; \log \rho (x_{n+1} |{\bf x}_n) \;   .
\label{h_nx}
\end{equation}
Equation (\ref{h_nx}) measures the information we have 
on $x_{n+1}$ if we know $x_1,..,x_n$
and (\ref{h_n}) is the average with respect to all possible 
sequences $x_1,..,x_n$.
If $h_n=0$ (the minimum possible) one has the maximum 
of information and next step can be predicted with certitude,
on the contrary, when $h_n$ attains its maximum,
no information is available. 

Since, more of the past it is known more it is the information,
the rate $h_n$ decreases when $n$ increases 
and the Shannon entropy $h=\lim_{n \to \infty} h_n$ 
measures the information we can extract from all the past
history. It should be noticed that
for a process which is $l$-step
Markovian  $h_n=h$ for all $n\ge l$.
In this case knowledge of more of the
last $l$ steps does not help to predict future. 

The Shannon entropy of the symbolic sequence associated to 
a dynamical system is known as Kolmogorov $\epsilon$-entropy~\cite{K}.
Eventually, by taking the supremum with respect to
all possible partitions it is possible to 
obtain the Kolmogorov-Sinai entropy,
which, in turn, equals the sum of positive Lyapunov exponents.

All above definitions deal with probabilities and 
with the idea that one has to consider many realizations of the process.
In practice, what one has to do is simply to consider a single
very long realization (much longer than $M^n$) of the process and 
to look for the frequency
of any of the sequences of length $n$.

The situation is less straightforward when
one deals with a random dynamical system

\begin{equation} 
y(t+1)=f(y(t), \omega(t+1))
\end{equation}
where  $\omega(t+1)$ is a random variable
which can be additive noise or a random parameter. 
In this case, the associated symbolic dynamics 
have random conditional probabilities.
In fact, $x(t+1)$ not only depends on the past history  
$..,x(1),x(2),..,x(t)$, but also on the present value $\omega(t+1)$
{\bf and} the past history $..,\omega(1),\omega(2),..,\omega(t)$ 
of a second stochastic process
which account for the uncontrolled random factors.
In this case we call the first subordinated process 
and the second fundamental process.
The reason is that for the problem we have in mind, the fundamental process 
is autonomous, which means that it has memory but 
$\omega(t+1)$ only depends on 
its own past history $..,\omega(1),\omega(2),..,\omega(t)$.
Nevertheless, all considerations which follows 
also holds for the most general case
in which the probability for $\omega(t+1)$ also
depends on the pasts history of the subordinated process
on $..,x(1),x(2),..,x(t-1)$.
In this more general case the distinction between fundamental and 
subordinated process is lost from a mathematical point of view,
although it will remain very significant from the information
point of view as we will see.

The problem is again to quantify the predictability
of the subordinated process, but
we are now ready to understand that
two different kind of situation may arise.
In the first the realization of the fundamental process is 
known (measurable or accessible information), 
in the second case it is unknown 
(non measurable or non accessible information).
Correspondingly, a different degree of
predictability on the subordinated process
is expected. More precisely, there is more information 
and predictability in the first case
(smaller entropy rate) and less in the second (larger entropy rate).

Before entering into the problem let us consider a much simpler
atemporal analogous which allows for clarifying the information context.
Consider a (subordinated) random variable $x$ whose probability
$\rho_\omega(x)$ depends on the actual value of a second 
(fundamental) random variable $\omega$.
One can think at this problem as if $\rho_\omega(x)$ 
is the conditional probability for $x$ given $\omega$.

The point we would like to focus can be better understood 
by the following example. 
Let us consider a coin toss in which the output is $x$,
but two different coins may be used.
The two coins are individuated by an index $\omega$
and they have different probabilities for $x$.
Given this simple situation, two different games may be played;
in both of them one has to guess the output of the toss
but the rule is different.
In the first game the player chooses at random the coin, gives a look 
at it, makes the guess and tosses;
in the second he makes the guess and only after  
he chooses the coin at random and tosses.

In the first game, at the moment of the guess,
$\omega$ is known and the information is measured by
the entropy
$-\sum_{ \{x\}}
 \rho_\omega(x) \; \log \rho_\omega(x)$.
Furthermore, since the coin has been chosen at random,
the entropy of the game before the start is the average
 
\begin{equation} 
\tilde{H}=
-\sum_{ \{x\}}
{<} \rho_\omega(x) \; \log \rho_\omega(x) {>}
\label{tildeH}
\end{equation}
One can easily check that $\tilde{H}= H_{x,\omega}-H_{\omega}$,
where $H_{x,\omega}$ is the entropy of the couple of random variables 
and $H_{\omega}$ is the entropy of $\omega$.

If the second game, $\omega$ is unknown at the moment of the guess,
and the utilized probability is the average
${<}\rho_\omega(x){>}$
Therefore, the information content is measured by

\begin{equation}
H=  - \sum_{ \{x\}}
 {<} \rho_\omega(x){>} \; \log {<}\rho_\omega(x){>}
\label{H}
\end{equation}

The inequalities 
$ \tilde{H} \le H \le H_{x,\omega}$ hold.
The first inequality trivially means that in the second game one has 
less information than in the first. The equality holds only in the
case of independence between the two random variables.
In turn, the second inequality becomes an equality only in case of complete
dependence of the two variables i.e. at given value $x$  
corresponds deterministically only a single value $\omega$.

The above discussion, although very simple, allows for the treatment 
of the intriguing case of stochastic process with random probability and,
therefore, also the case of dynamical systems with noise.

In the first scenario, once probabilistically generated,
the sequence $..,\omega(1),..,\omega(t),..$ can be 
measured and it can be
treated as a known ordinary time dependent 
function. In consequence, the average entropy of a sequence 
of length $n$ is given by
\begin{equation}
\tilde{H_n}
= - \sum_{ \{x_1,..,x_n\}}
{<} \rho_\omega ({\bf x}_n) \;
\log  \rho_\omega ({\bf x}_n){>}
\end{equation}
since $\rho_\omega ({\bf x}_n)$ is the 
probability for the sequence $x_1,..,x_n$ given 
the realization of the fundamental process. 

The entropy rate
$\tilde{h}_n=\tilde{H}_{n+1}-\tilde{H}_n$
represents now the average information one has 
on $x(t+1)$ if one knows its previous $n$ steps $x(1),..,x(t)$
and also one knows {\bf all} the 
fundamental process from the most recent $\omega(t+1)$
to the more remote past. 

In this case, in analogy with the atemporal example,
one can easily show that the Shannon entropy 
$ \tilde{h} = \lim_{n\to\infty} \tilde{h}_n$ is the difference between
the Shannon entropy associated with the couple process  
and that the Shannon entropy of the fundamental process
i.e. $ \tilde{h}=h_{x,\omega}-h_\omega$.
Therefore, from a practical point of view, what one has to do is to
generate or to consider a very long sequence of both the processes and 
then measure the frequency of the sequences 
$(x_1; \, \omega_1),..,(x_n; \, \omega_n),..$ and,
separately,  of the sequence $\omega_1,..,\omega_n,..$ .

Furthermore, $\tilde{h}$ is nothing else
that the characterization of complexity which refers to the separation
of two nearby trajectories corresponding to the same realization of
the fundamental process (noise).
Notice, in fact, that one also trivially has
$h=\lim_{n \to \infty} \tilde{H}_n/n$.
Since $\tilde{H}_n/n$ is an average 
of a quantity which is non random in the limit
one can replace the average by simply considering 
the typical value, i.e. that value corresponding to a single realization
of the fundamental process.
This realization only plays the role of an ordinary
function of time.
The Pesin relation assures, in this case, that 
$\tilde{h}$ equals the sum of positive
Lyapunov exponents corresponding to the separation
of trajectories under the
same realization of the fundamental process.
This quantity is often taken as a measure of
complexity for chaotic noisy systems.
Nevertheless, let us stress once again 
that it is a measure of predictability
{\bf only if} the noise realization itself can be measured
with infinite precision.

Let us consider the more realistic case
in which on the contrary $\omega(t)$ cannot be measured.
In a way which is completely analogous to 
(\ref{H}) one finds that the entropy 
associated to a sequence of length $n$ is

\begin{equation}
H_n
= - \sum_{ \{x_1,..,x_n\}}
{<} \rho_\omega (x){>} \;
\log {<} \rho_\omega(x){>}
\end{equation}
and the entropy rate $h_n=H_{n+1}-H_n$ describes
now the information content of 
$n$ steps of the subordinated process when the fundamental
process is unknown. The Shannon entropy  $h =\lim_{n \to \infty}
h_n$ is the maximum information which is available from
the past.  

In practice, in order to obtain $h_n$, one has to generate the long sequence 
$..,x(1),x(2),..,x(t),..$ and measure the frequency  
of the sequences without regarding at the fundamental process.

We stress that this characterization of
predictability may differ a lot from the first one.
For example, even non chaotic dynamical systems with negative
or vanishing Lyapunov exponents ($\tilde{h}=0$) may have a positive
Shannon entropy $h$ and to be largely unpredictable.
More generally the inequality  $\tilde{h} \le h \le h_{x,\omega}$
holds. The qualitative 
understanding of this inequality is straightforward 
since $\tilde{h}$ refers to the
case in which the  additional information on the fundamental 
process is available.
On the other side, it is easy to check that $h$ is smaller or equal than 
the entropy of the couple process. Since 
$\tilde{h} =h_{x,\omega}- h_x$ one can also rewrite
$ h_{x,\omega}- h_x \le h \le h_{x,\omega}$.
The first inequality becomes an equality only when
the two process are reciprocally independent,
while the the second one when they are deterministically
linked. 

Unfortunately, while one has that $\tilde{h}$ can be computed by 
means of the Lyapunov exponents of a typical trajectory, 
there is not a corresponding general recipe for $h$.
Nevertheless, a very sensitive approximate approach
based on the use of the rate of separation of
two nearby trajectories which correspond 
to {\bf two} different realization of the fundamental
has been recently proposed~\cite{PSV,LPV,LSV,LPPV} .

Let us now give a simple example which may help to understand
an important point: a subordinated process which is 
memoryless when the realization of the fundamental process is given,
may have a long memory when, on the contrary, 
the fundamental process is unknown.
This is because part of the information carried by
the fundamental is encoded in the history of the subordinated.
The long memory of the subordinated 
may be induced even if the  fundamental itself is Markov. 
The more the fundamental is correlated, longer is 
the memory of the subordinated.
As a consequence, the difference between $h$ and $\tilde{h}$
decreases when the correlation of the fundamental
increases and it disappears in case this correlation is complete.

Let us consider, as an example, the following simple dynamical system
\begin{equation}
y(t+1) =2^{\omega(t+1)} y(t) \;\;\;\;\;\;\;\;\;\;\;\;\;\;\;\;
\;\;\;\;\;\;  mod.\;\;\;\; 1
\end{equation}
where the $\omega(t)$ are Markovian variables which can take
the two possible values $0,1$ with equal probability
and which persist in their value with probability $p$.

Now let us consider the most trivial partition of
the accessible phase space: the two intervals $[0,1/2)$, $[1/2,1)$
identified respectively by the symbols $x=0$ and $x=1$.
The symbolic dynamics is very easy to construct:
when  $\omega(t+1)=0$  then  $x(t+1)=x(t)$,  when  $\omega(t+1)=1$ 
then  $x(t+1)=0,1$  with same probability  $1/2$.

Than it is straightforward to obtain, independently on $p$ 
the entropy rate $\tilde{h}_n=\frac{1}{2}\log 2$ 
for $n \ge 1$. This entropy equals 
the typical rate of separation of the trajectories.

On the contrary $h_n$ strongly depends 
on $p$. If $p=0$ than $ h_n=-\frac{3}{4} 
\log \frac{3}{4}-\frac{1}{4}\log \frac{1}{4}$
for  $n \ge 1$.
This result can be easily interpreted in this way:
the averaged probability that $x(t)$ 
persist in its value is $3/4$, while the averaged probability for a 
change is $1/4$. 

In fig 1. $h_n$ is plotted for three different values of $p\neq0$
and compared both $\tilde{h}_n$ and the value of $h_n$
corresponding to $p=0$.
When $p\neq0$, one see from fig 1. that $h_n$ is strictly monotone
and converges to $h$ in more than a single step. 
This behavior implies that in absence of informations 
on the fundamental the subordinated is not Markovian even 
if the couple process itself is Markovian.
This behavior also implies that when $n$ increases, 
the knowledge of the subordinated trajectory 
can be used to better forecast. 

Furthermore, one observes in fig. 1 that the 
asymptotic difference $h-\tilde{h}$ reduces when $p$ increases
and almost vanishes for very large $p$.
This is of easy understanding since, in practice,
one guesses the value of $x(t+1)$ 
by looking at the length of the previous sequence   
of symbols with same persistent value. 
The larger is $p$, the longer are  
these sequences which carry information.
  
In other words, since the more it is possible to use of the past, 
the more the information is recovered,  
the difference $h-\tilde{h}$ reduces when $p$ increases.  

\begin{figure}
\centerline{\psfig{file=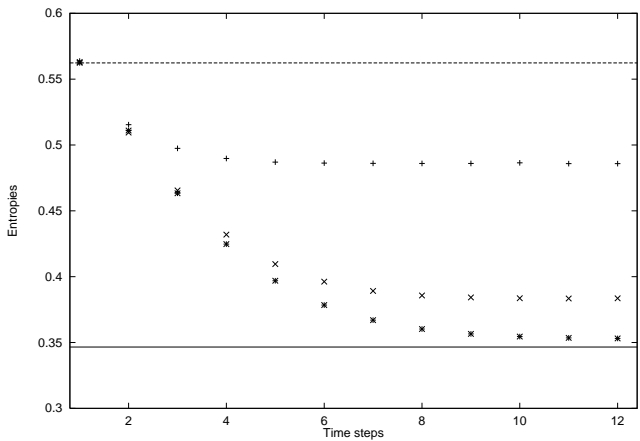,width=8.5cm,angle=270}} 
\caption{The entropy rate $h_n$ is plotted for three different values of $p$:
$p=.9$ (crosses), $p=.99$ (slanting crosses),$p=.999$ (asterisks).
For comparison, both $\tilde{h}_n$ (line) and the value of $h_n$
corresponding to $p=0$ (dots) are reported.
} 
\label{fig1}
\end{figure} 

This example shows that the topic of this paper may be quite relevant 
when one deal with historical data, i.e. single non reproducible 
sequences of symbols, where the joint probabilities 
depend on some stochastic parameters. For example this phenomenology 
is typical of finance where the random parameters
reflect economic factors which may be unknown to a given investor.   
Investors which trust in fundamental analysis 
strongly believe that all information 
is reflected in the price dynamics.
The previous example also shows the limit of this point of view.
In fact, the missing information is only partially reflected by data
and only information on longly persistent macro-economic factors 
can be totally recovered.

Let us resume the result of this paper as follows:

\noindent
if the fundamental process is known, the predictability is measured
by $\tilde{h}$ which can be computed in practice as
the Shannon entropy of a single long realization of the couple process 
minus the Shannon entropy of a single long realization 
of the fundamental process. 
This entropy equals the sum of positive Lyapunov 
exponents associated to the separation of nearby trajectories 
under the same realization of the noise;

\noindent
if the fundamental process is unknown, the predictability is measured
by $h$ which can be computed in practice as
the Shannon entropy of a single long 
realization of the subordinated process.
Unfortunately, there is not an exact recipe 
which allows for the calculation of $h$ by means of Lyapunov exponents
and some more refined approximate techniques have 
to be used~\cite{PSV,LPV,LSV,LPPV}.

\bigskip
The author is indebted with Angelo Vulpiani for an infinite number of
discussion on complexity, chaos and predictability. He
also thanks Roberto Baviera, Michele Pasquini and Yi-Cheng Zhang 
for discussion and suggestions concerning the topic of this work.
This research has been partially supported by the European Network 
contract FMRXCT980183.


\begin{thebibliography}{99}

\bibitem{CFH} J.P. Crutchfield, D.J. Farmer and B.A. Huberman,
Phys. Rep {\bf 92}, (1982), 45.

\bibitem{BJS} A.R. Bulsara, E.W. Jacobs and W.C. Schieve,
Phys. Rev. A {\bf 42}, (1990), 5837

\bibitem{MT} K. Matsumoto and I. Tsuda,
J. Stat. Phys. {\bf 31}, (1983), 87

\bibitem{PSV} G. Paladin, M. Serva and A. Vulpiani, 
Phys. Rev. Lett. {\bf 74}, (1995), 66.

\bibitem{LPV} V. Loreto, G. Paladin and A. Vulpiani, 
Phys. Rev. E {\bf 53}, (1996), 2087.

\bibitem{LSV} V. Loreto, M. Serva and A. Vulpiani, 
Proceedings of {\it Trends in complexity}, (Yokohama 1996).
Int. J. Mod. Phys. B {\bf 12}, (1998), 225.

\bibitem{LPPV} V. Loreto, G. Paladin, M. Pasquini and A. Vulpiani,
Physica A {\bf 232}, (1996), 189.

\bibitem{SW} C.E. Shannon and W. Weaver,
{\it The mathematical theory of communications},
Univ. of Illinois Press, Urbana (1949).

\bibitem{K} A.N. Kolmogorov,
IRE Trans. Inf. Theory {\bf 1}, (1956), 102.


\end{thebibliography}
\end{document}